\documentclass[12pt]{article}

\usepackage{graphicx}
\usepackage{amsfonts}
\newtheorem{proposition}{Proposition}

\newtheorem{definition}[proposition]{Definition}

\def\+{{+\!\!\!+}}

\def\d{\partial}

\def\pmb#1{\setbox0=\hbox{#1}%
\kern.0em\copy0\kern-\wd0 
\kern-.04em\copy0\kern-\wd0 
\kern.08em\copy0\kern-\wd0 
\kern-.04em\raise.0433em\box0 }         
 
\newcommand{\nc}{\newcommand} 
\nc{\beq}{\begin{equation}} 
\nc{\eeq}[1]{\label{#1}\end{equation}} 
\nc{\ber}{\begin{eqnarray}} 
\nc{\eer}[1]{\label{#1}\end{eqnarray}} 
\nc{\pek}[1]{\cite{#1}} 
\nc{\enr}[1]{(\ref{#1})} 
\nc{\kal}[1]{{\cal{#1}}} 
\nc{\dott}{\;\cdot\;} 
\newcommand{\Section}[1]{\section{#1} \setcounter{equation}{0}}

\def\0 {\nonumber}

\begin{document} 
\setcounter{page}{0}
\newcommand{\inv}[1]{{#1}^{-1}} 
\renewcommand{\theequation}{\thesection.\arabic{equation}} 
\newcommand{\be}{\begin{equation}} 
\newcommand{\ee}{\end{equation}} 
\newcommand{\bea}{\begin{eqnarray}} 
\newcommand{\eea}{\end{eqnarray}} 
\newcommand{\re}[1]{(\ref{#1})} 
\newcommand{\qv}{\quad ,} 
\newcommand{\qp}{\quad .} 
\begin{titlepage} 
\begin{center} 

\hfill SISSA 64/2005/FM\\  
\hfill UUITP-17/05\\

\vskip .3in \noindent 


{\Large \bf{On topological M-theory}} \\

\vskip .2in 

{\bf Giulio Bonelli$^{a}$}, {\bf Alessandro Tanzini$^a$} 
and {\bf Maxim Zabzine$^{b,c}$,}

\vskip .05in 
$^a${\em\small International School of Advanced Studies (SISSA) and INFN, Sezione di Trieste\\
 via Beirut 2-4, 34014 Trieste, Italy} 
\vskip .05in 
$^b${\em School of Mathematical Sciences, Queen Mary, University of London, \\
Mile End Road, London, E1 4NS, UK}
\vskip .05in 
$^c${\em  Department of Theoretical Physics, \\
Uppsala University,
Box 803, SE-751 08 Uppsala, Sweden }
\vskip .5in
\end{center} 
\begin{center} {\bf ABSTRACT }  
\end{center} 
\begin{quotation}\noindent  
 We construct a gauge fixed action for topological membranes on $G_2$-manifolds such that
 its bosonic part  
 is the standard membrane theory in a particular gauge.
 We prove that quantum mechanically
the path-integral in this gauge localizes on associative submanifolds. 
Moreover on $M\times S^1$ the 
   theory naturally reduces to the standard A-model on Calabi-Yau manifold and 
    to a membrane theory localized on special Lagrangian submanifolds.
We discuss some properties of topological membrane theory on 
     $G_2$-manifolds. 
We also generalize our construction to topological $p$--branes on special manifolds
by exploring a relation between vector cross product structures and TFTs.
 \end{quotation} 
\vfill 
\eject 


\end{titlepage}

\section{Introduction}

The notion of topological M-theory has been introduced in \cite{Dijkgraaf:2004te} (also 
 for earlier proposal see \cite{Gerasimov:2004yx})
 as unifying description of the topological A- and B-models. This is very much in analogy
 with the connection between the physical superstring and and the physical M-theory. 
  In \cite{Dijkgraaf:2004te} the analysis has been done at the classical level
  of the ``effective'' actions. Different arguments in favor of topological 
 M-theory have been proposed in \cite{Grassi:2004xr,Baulieu:2004pv,Nekrasov:2004vv}. 

 In this note we propose a microscopic description of topological M-theory on 
 seven dimensional $G_2$-manifold as a topological membrane theory.  
 Namely we construct the gauge fixed action $S_{GF}$ for the topological membrane 
\beq
 S_{top} = \int \,\,X^*(\Phi),
\eeq{topmebranebegin}
 where $\Phi$ is a closed three form associated with a $G_2$-structure. 
 The bosonic part of $S_{GF}$ turns out to be the standard membrane theory in a particular gauge. 
 Moreover on $CY_6\times S^1$ the action $S_{GF}$ naturally reduces to A-model. 
 The proposed membrane theory is localized on associative cycles.
 It is well-known that membrane instantons on $G_2$-manifold are  
 given by associative three submanifolds \cite{Becker:1995kb}.
   The contribution of membrane instantons on $G_2$-manifold to the superpotential
   of ${\cal N}=1$ compactifications of M-theory have been studied in
    \cite{Harvey:1999as} and \cite{Beasley:2003fx}. 
 
Actually, in the present work we do not couple the topological  membrane model
to $3D$ gravity on the world-volume and therefore a full comparison with the topological 
string can not be performed yet. 
We plan to discuss this problem in a separate work. 

 Recently using the  Mathai-Quillen formalism the authors \cite{Anguelova:2005cv}
  proposed a gauge fixed action for the topological membrane.
 However the bosonic part of this model 
 is unusually highly polynomial and this obscures the relation to the usual membrane theory, since in \cite{Anguelova:2005cv} agreement is found
only up to quadratic order.
Moreover, 
the relation with A-model is shown at the level of zero section
and not for the gauge fixed action. 
 Further we will comment more 
 on the relation between our construction and the one proposed in \cite{Anguelova:2005cv}.
 In \cite{Bao:2005yx} the authors discuss 
 topological membranes using the Green-Schwarz formalism. 
 
The structure of the paper is as follows. 
In Section \ref{membrane} we recall the description 
 of membrane instantons on manifolds with $G_2$-structure. On a $G_2$-manifold the instantons 
 correspond to three dimensional associative submanifolds. 
In Section \ref{gfaction} we construct 
 the gauge fixed action for the topological membrane (\ref{topmebranebegin}). In our treatment 
 we follow closely the lagrangean approach advocated by  Baulieu and Singer 
in \cite{Baulieu:1989rs} for the topological sigma model.
Its extension to our membrane theory proves that 
the path-integral evaluation in our gauge
is localized on membrane instantons, that is on associative submanifolds. 
We discuss also further properties of the gauge fixed action. 
In the next Section we go through the Hamiltonian treatment and 
 make a few comments regarding the previous work \cite{Bonelli:2005ti}.  
  Section \ref{observables} is devoted to the discussion of observables and moduli spaces.
  In Section \ref{vcp} we collect some observations about the general relations between 
   vector product structures and TFTs generalizing our construction.
   The Appendices collect some relevant properties of $G_2$-manifolds and vector 
    cross product structures.

\section{Membrane instantons on $G_2$-manifolds}
\label{membrane}

 In this Section we present a natural and elementary approach to the  classical aspects
  of membrane instantons.
 
  Let us consider the Euclidean membrane theory defined by 
   the following Nambu-Goto action
   \beq
    S = \int d^3\sigma\,\,\sqrt{\det(\d_\alpha X^\mu g_{\mu\nu} \d_\beta X^\nu)},
   \eeq{NGaction}
 where $\alpha =0,1,2$. In (\ref{NGaction}) we have chosen units such 
 that the membrane tension is one. 
  Introducing 
  the auxiliary world-volume metric $h_{\alpha\beta}$ the action (\ref{NGaction}) can be obtained as
  the stationary value of
  \beq
   S = \frac{1}{2} \int d^3\sigma\,\,\sqrt{h} (h^{\alpha\beta} \d_\alpha X^\mu 
    g_{\mu\nu} \d_\beta X^\nu  -1) .
  \eeq{Polaction}
  under arbitrary variations of $h_{\alpha\beta}$.
 Let us now fix the gauge symmetry. 
Unlike the string case, there is not enough gauge symmetry 
  to fix the whole auxiliary metric $h_{\alpha\beta}$ which has six independent components.
   However using reparameterization symmetry we can fix the components $h_{0\beta}$ to be
   \beq
    h_{0a} = 0,\,\,\,\,\,\,\,\,\,\,\,\,\,\,\,
    h_{00}= \det \left(h_{ab}\right)
  \eeq{gaugefix}
 where $h_{ab}$, with $a,b = 1,2$, are the remaining spatial components 
of the auxiliary metric. 
Once we have chosen this gauge, no further components of 
  $h_{\alpha\beta}$ can be fixed. Globally this gauge can be only chosen when the membrane 
   world-volume is of the form $\Sigma_2 \times S^1$ (also
     $S^1$ can be replaced by either an interval or a real line) with $\Sigma_2$ being a Riemann
      surface. Therefore this lagrangean approach to gauge fixing is equivalent to the usual hamiltonian one.
 After fixing the stationary condition for the remaining $h_{ab}$ 
      in this gauge the membrane action becomes
      \beq
      S = \frac{1}{2} \int d^3\sigma\,\,\left ( \dot{X}^\mu g_{\mu\nu} \dot{X}^\nu +
       \det (\d_a X^\mu g_{\mu\nu} \d_b X^\nu) \right ),
      \eeq{gaugefixacbos}  
    where $\dot{X}^\mu \equiv \d_0 X^\mu$, which is the well known gauge fixed
     Euclidean membrane action on the Riemannian manifold $(M,g)$.
      
 Now consider the following bound
\beq
\int d^3\sigma\,\, \left ( \dot{X}^\mu \pm \Phi^\mu_{\,\,\,\nu\rho} \d_1 X^\nu \d_2 X^\rho \right ) g_{\mu\lambda}
 \left ( \dot{X}^\lambda \pm \Phi^\lambda_{\,\,\,\sigma\tau}\d_1 X^\sigma \d_2 X^\tau
  \right ) \geq 0 ,
\eeq{boudnao20}      
 where $\Phi_{\mu\nu\rho}$ is a $3-$form and $g_{\mu\nu}$ a Riemannian metric on $M$.
  If $M$ is a seven dimensional manifold with $G_2$ structure given by $\Phi$ and $g$
via the vector cross product relation (see Appendix B),
  then the bound (\ref{boudnao20}) can be rewritten as follows
\beq
 \frac{1}{2} \int d^3\sigma\,\,\left ( \dot{X}^\mu g_{\mu\nu} \dot{X}^\nu +
       \det (\d_a X^\mu g_{\mu\nu} \d_b X^\nu) \right ) \geq \mp \frac{1}{6} \int X^*(\Phi).
\eeq{membrabound1}
  If $\Phi$ is a closed form then the term on the right hand side of (\ref{membrabound1})
   is topological. Thus for the manifold with $G_2$-structure $(M,g,\Phi)$ such 
    that $d\Phi =0$ we obtain the following membrane instantons
    \beq
    \dot{X}^\mu \pm \Phi^\mu_{\,\,\,\nu\rho} \d_1 X^\nu \d_2 X^\rho = 0,
    \eeq{definosnal}
     which minimize the Euclidean action (\ref{gaugefixacbos}).
We call a map $X$ from $\Sigma_3$ to $M$ which satisfies (\ref{definosnal}) 
an associative map.

Actually, we can show that the condition 
(\ref{definosnal}) is equivalent to the calibration condition
\beq
 d\, vol (\Sigma_3) = \mp \frac{1}{6}  \,X^*(\Phi),
\eeq{calcbrajwpw}
where $d\, vol$ is the volume element induced by $g$ and the pull-back is along
the associative map. In fact by multiplying (\ref{definosnal}) by 
$g_{\mu\nu}\partial_\alpha X^\nu$ we get 
$ h_{0a}= g_{\mu\nu} \partial_a X^\mu \dot X^\nu = 0$ and
$(\dot X)^2 \pm \frac{1}{6} X^*(\Phi) = 0$. On the other hand,
by squaring (\ref{definosnal}) and eliminating $X^*(\Phi)$
by the previous equations we get for the induced metric
$h_{00}= {\rm det} (h_{ab})$.
Hence
\beq
dvol (\Sigma_3) = \sqrt{h} = h_{00} = \mp \frac{1}{6} X^*(\Phi)
\eeq{calibre}

If we require the manifold $M$ to be of $G_2$-holonomy\footnote{From now on we refer 
 to the manifolds of $G_2$-holonomy as $G_2$-manifolds.} (i.e. either
 $\nabla_\mu \Phi_{\nu\rho\sigma}=0$ or $d\Phi=0$ and $d*\Phi=0$) then the instantons (\ref{definosnal})
are interpreted as associative submanifolds of $M$, namely 
submanifolds calibrated by $\Phi$ \cite{calib}.

 Next consider $M_6$ to be a Calabi-Yau threefold with K\"ahler form
 $\omega$ and holomorphic three form $\Omega$. Then there is a natural $G_2$-structure 
 $\Phi$ on the product $M_7 = M_6 \times S^1$ given by 
  \beq
   \Phi = Re\, \Omega + dX^7 \wedge \omega ,
  \eeq{instajdpel}
   where we choose the coordinates $\mu = (N, 7)$ with the
 uppercase Latin letters denoting the coordinates along $M_6$ and $X^7$ a coordinate along
 $S^1$. This induces the product metric on $M_6 \times S^1$, with the flat metric on $S^1$
\beq
 g_{\mu\nu} = \left ( \begin{array}{ll}
 g_{MN} & 0 \\
   \,\,\, 0 & 1
   \end{array} \right ).
\eeq{metrifcnso}
 With the appropriate orientation on $M_6 \times S^1$ we have
\beq
 *\Phi = - d X^7 \wedge Im\,\Omega + \frac{1}{2} \omega^2. 
\eeq{metriapo3}
 Thus $M_6\times S^1$ is a $G_2$-manifold.
 On $M_6 \times S^1$ the associative 3-cycles wrapping $S^1$
are of the form $\Sigma_3 = \Sigma_2 \times S^1$, where $\Sigma_2$ is an 
(anti)holomorphic 
 curve in $M_6$, while the associative 3-cycles localized along $S^1$
 correspond to special Lagrangian submanifolds with phase zero. 
 Correspondingly 
 the Euclidean action (\ref{gaugefixacbos}) reduces on $M_6$ 
  either to string theory in conformal gauge
\beq
 S = \frac{k}{2} \int d^2\sigma \,\,\left ( \dot{X}^N g_{NM} \dot{X}^M +
       \d_1 X^N g_{NM} \d_1 X^M) \right )
\eeq{stringaction}
where $k$ is the $S^1$ winding, 
 or to membrane theory in the same type of gauge as before
\beq
 S = \frac{1}{2} \int d^3\sigma\,\,\left ( \dot{X}^N g_{NM} \dot{X}^M +
  \det (\d_a X^N g_{NM} \d_b X^M) \right ). 
\eeq{membraact3cy}
These reductions can also be done at the level of membrane instantons
(\ref{definosnal}). Thus the holomorphic curves (calibrated
 by $\omega$) are the instantons for (\ref{stringaction}) and the special 
 Lagrangian submanifolds with phase zero (calibrated by 
 $Re\, \Omega$) are instantons for (\ref{membraact3cy}). However the special Lagrangian
 submanifolds with phase zero are not the most general instantons 
 for the action (\ref{membraact3cy})
-- see Section \ref{vcp} for further discussion. 

Indeed there is a family of $G_2$-structures on $M_6 \times S^1$ \cite{karigiannis}
\beq
 \Phi_\theta = Re\,(e^{i\theta}\Omega) + dX^7 \wedge \omega,
\eeq{familyG2str}
 where now the Calabi-Yau structure on $M_6$ is given by $e^{i\theta}\Omega$ and $\omega$. It is well-known 
 that one can change the holomorphic form $\Omega$ by a multiplicative phase while preserving the Ricci-flat 
 metric. With this new $G_2$-structure one can repeat the same considerations as above. However now 
 the associative manifolds localized in $S^1$ correspond to special Lagrangian submanifolds of $M_6$ 
 with the phase $\theta$, i.e. calibrated by $Re\,(e^{i\theta}\Omega)$. 

\section{The gauged fixed action}
\label{gfaction}

In this Section we consider the gauge fixing for the following topological 
 membrane theory
 \beq
  S = - \frac{1}{6}\int\limits_{\Sigma_3} \,\,X^*(\Phi),
 \eeq{topaction}
 where $\Phi$ is the three form associated to a $G_2$-structure on a seven dimensional manifold $M_7$. 
 We assume that $\Phi$ is closed. In our treatment 
  we closely follow Baulieu and Singer  \cite{Baulieu:1989rs} generalizing their method to membranes.
 
  The gauge symmetry of the action is 
  \beq
   \delta X^\mu = \epsilon^\mu .
  \eeq{gaugetransf}
  The corresponding BRST operator $s$ is  defined as follows  
  \beq
  sX^\mu = \psi^\mu,\,\,\,\,\,\,\,\,\,\,
  s\psi^\mu =0,\,\,\,\,\,\,\,\,\,\,
  s\bar{\psi}^\mu = b^\mu,\,\,\,\,\,\,\,\,\,\,
  sb^\mu =0,
  \eeq{BRST1}
   where $\psi^\mu$ is the ghost associated to $\epsilon^\mu$. The ghost numbers 
    are respectively $0,1,-1,0$ for $X^\mu, \psi^\mu, \bar{\psi}^\mu, b^\mu$.
   We will now fix the symmetry (\ref{gaugetransf}) and obtain a gauged fixed action
   which is quadratic in the velocities. We choose the following gauge function
\beq
 {\cal F}^\mu = \dot{X}^\mu + \Phi^\mu_{\,\,\,\nu\rho} \d_1 X^\nu \d_2 X^\rho 
+ \frac{1}{2}\Gamma^\mu_{\,\,\,\sigma\rho}
 \bar{\psi}^\sigma \psi^\rho,
\eeq{gaugefunct}
 where $\Gamma_{\mu\nu\rho} = 1/2(g_{\mu\nu,\rho} + g_{\mu\rho,\nu} - g_{\nu\rho,\mu})$
 with $g_{\mu\nu,\rho} \equiv \d_\rho g_{\mu\nu}$. 
The quadratic term in 
 the ghosts
 in (\ref{gaugefunct}) is necessary for manifest general covariance. 
   The BRST invariant gauge fixed action is obtained by adding to the classical action (\ref{topaction}) an $s$-exact  gauge fixing term and reads  
    \beq
     S_{GF} =-\frac{1}{6} \int X^*(\Phi) + \int d^3\sigma\,\, s \left ( \bar{\psi}^\mu (g_{\mu\nu} \dot{X}^\nu 
      + \Phi_{\mu\nu\rho} \d_1 X^\nu \d_2 X^\rho + \frac{1}{2} \Gamma_{\mu\sigma\rho}
       \bar{\psi}^\sigma \psi^\rho - \frac{1}{2} g_{\mu\nu} b^\nu ) \right ) .
    \eeq{startgaugefixed}
 Using the definition (\ref{BRST1}) and eliminating $b^\mu$ by its algebraic equation
  of motion
  \beq
  b^\mu = \dot{X}^\mu + \Phi^\mu_{\,\,\,\nu\rho} \d_1 X^\nu \d_2 X^\rho + \Gamma^\mu_{\,\,\,\sigma\rho}
   \bar{\psi}^\sigma \psi^\rho
   \eeq{eqforb}
   we arrive to the following gauged fixed action
   $$     S_{GF}=\int d^3\sigma \left ( \frac{1}{2} \dot{X}^\mu g_{\mu\nu} \dot{X}^\nu
  + \frac{1}{2}\det (\d_a X^\mu g_{\mu\nu} \d_b X^\nu)
     - \bar{\psi}^\mu g_{\mu\nu} \nabla_0 \psi^\nu \right . $$
\beq
\left . -\Phi_{\mu\nu\rho} \bar{\psi}^\mu \nabla_a \psi^\nu
      \d_b X^\rho \epsilon^{ab} 
      -\frac{1}{2} \epsilon^{ab} \d_a X^\nu \d_b X^\rho \bar{\psi}^\mu \psi^\lambda
      \nabla_\lambda \Phi_{\mu\nu\rho}
      + \frac{1}{4} {\cal R}_{\mu\sigma\lambda\rho} \bar{\psi}^\mu
      \psi^\rho \bar{\psi}^\sigma \psi^\lambda \right ),
   \eeq{fullactionforclos}
  where
  \beq
  \nabla_\alpha \psi^\mu = \d_\alpha \psi^\mu  + \Gamma^\mu_{\,\,\,\rho\lambda} \d_\alpha X^\rho
   \psi^\lambda
  \eeq{nbladef}
 and
\beq
 {\cal R}^\mu_{\,\,\,\sigma\rho\lambda} = \Gamma^\mu_{\,\,\,\lambda\sigma,\rho} - \Gamma^{\mu}_{\,\,\,\rho\sigma,\lambda}
 + \Gamma^\mu_{\,\,\,\rho\tau} \Gamma^\tau_{\,\,\,\lambda\sigma} - 
 \Gamma^\mu_{\,\,\,\lambda\tau} \Gamma^\tau_{\,\,\,\rho\sigma} .
\eeq{defincurvature}
For sake of simplicity,
from now on we assume that $\Phi$ is also coclosed, i.e. $\nabla_\lambda \Phi_{\mu\nu\rho}=0$
   and thus the manifold is of $G_2$-holonomy.
 Assuming $G_2$-holonomy the gauge fixed action (\ref{fullactionforclos}) becomes  
$$     S_{GF}=\int d^3\sigma \left ( \frac{1}{2} \dot{X}^\mu g_{\mu\nu} \dot{X}^\nu
  + \frac{1}{2}\det (\d_a X^\mu g_{\mu\nu} \d_b X^\nu)
     - \bar{\psi}^\mu g_{\mu\nu} \nabla_0 \psi^\nu  \right . $$
\beq
\left . -\Phi_{\mu\nu\rho} \bar{\psi}^\mu \nabla_a \psi^\nu
      \d_b X^\rho \epsilon^{ab} + \frac{1}{4} {\cal R}_{\mu\sigma\lambda\rho} \bar{\psi}^\mu
      \psi^\rho \bar{\psi}^\sigma \psi^\lambda \right ).
   \eeq{fullaction}
  The action (\ref{fullaction}) is invariant under the following BRST symmetry
    \beq
  sX^\mu = \psi^\mu,\,\,\,\,\,\,\,\,\,\,
  s\psi^\mu =0,\,\,\,\,\,\,\,\,\,\,
  s\bar{\psi}^\mu =  \dot{X}^\mu + \Phi^\mu_{\,\,\,\nu\rho} \d_1 X^\nu \d_2 X^\rho + \Gamma^\mu_{\,\,\,\sigma\rho}
   \bar{\psi}^\sigma \psi^\rho ,
  \eeq{BRST2}
  which is nilpotent on-shell only unlike (\ref{BRST1}). The action (\ref{fullaction}) can be rewritten 
   as follows
  \beq
   S_{GF} = -\frac{1}{6} \int X^*(\Phi) + \frac{1}{2}
   \int d^3\sigma\,\, s \left ( \bar{\psi}^\mu (g_{\mu\nu} \dot{X}^\nu 
      + \Phi_{\mu\nu\rho} \d_1 X^\nu \d_2 X^\rho)\right ).
  \eeq{localpropsioe}
 Thus due to standard arguments (e.g., see \cite{Witten:1991zz}) 
 the model is localized on the solutions of (\ref{definosnal}) which correspond to
 associative three manifolds. In (\ref{localpropsioe}) the topological term depends only on the cohomology class
  of  $\Phi$ and the homotopy class of the map $X$. 

If we assume that $H^3(M_7, \mathbb{Z}) =
   \mathbb{Z}$, we can normalize $\Phi$ such that the periods
    of $\frac{1}{6}\Phi$ are integer multiples of $2\pi$,  that is
    \beq
    \frac{1}{6}\int\limits_{\Sigma_3} X^*(\Phi) = 2\pi n,
    \,\,\,\,\,\,\,\,\,\,\,n \in \mathbb{Z}
    \eeq{degreenorm}
where $n$ is the instanton number for the associative map (\ref{definosnal}). 
Therefore the path integral is reduced to a sum of the integrals over the 
moduli space ${\cal M}_n$ of associative maps of degree $n$.
Actually (\ref{degreenorm}) is not well defined at the quantum
mechanical level due to a parity anomaly \cite{Atiyah:2001qf}
(see Sect.5 for further discussions).
The amplitudes of our topological theory 
do not depend on the way we describe the associative 
maps. Namely in (\ref{definosnal}) we choose a specific 
coordinate with distinguished direction $\alpha =0$. 
However any variation with respect to this choice appears in the path
integral as a BRST--exact term.
Thus indeed our specific parametrization of associative maps is irrelevant 
for the theory.

 The action (\ref{fullaction}) has another set of BRST transformations
  \beq
  \bar{s}X^\mu = \bar{\psi}^\mu,\,\,\,\,\,\,\,\,\,\,
  \bar{s}\bar{\psi}^\mu =0,\,\,\,\,\,\,\,\,\,\,
  \bar{s} \psi^\mu =  \dot{X}^\mu - \Phi^\mu_{\,\,\,\nu\rho} \d_1 X^\nu \d_2 X^\rho - \Gamma^\mu_{\,\,\,\sigma\rho}
   \bar{\psi}^\sigma \psi^\rho ,
  \eeq{BRST2extra}
 which are nilpotent only on-shell. The two BRST transformations above form the following on-shell algebra
\beq
 s \bar{s} + \bar{s} s = 2 \d_0.
\eeq{onshelwyqo}
 We will comment more on the transformations (\ref{BRST2}) and (\ref{BRST2extra}) in Section \ref{ham}. 
  
 Furthermore we can consider the theory (\ref{fullaction}) with its BRST 
 symmetry (\ref{BRST2})
 on $M_7 = M_6 \times S^1$ on which 
 we assume the structure (\ref{instajdpel})-(\ref{metriapo3}). 
 There are two interesting sectors:
 the configurations which wrap $S^1$ and the configurations which are localized on $S^1$. First let us 
 consider the configurations which wrap $S^1$. Assuming that $X^7 = k \sigma_2$ with $\psi^7=\bar{\psi}^7=0$
 and the other fields independent on $\sigma_2$, the action (\ref{fullaction}) reduces to 
$$     S_{GF}=
 k \int d^2\sigma \left ( \frac{1}{2} \dot{X}^N g_{NM} \dot{X}^M + \frac{1}{2} \d_1 X^N g_{NM} \d_1 X^M
     - \bar{\psi}^N g_{NM} \nabla_0 \psi^M  \right .$$ 
\beq 
\left . -\omega_{MN} \bar{\psi}^M \nabla_1 \psi^N
     + \frac{1}{4} {\cal R}_{MNPL} \bar{\psi}^M
      \psi^L \bar{\psi}^N \psi^P \right ),
   \eeq{fullaction1}
 where we have used (\ref{instajdpel}) and (\ref{metrifcnso}).
   In its turn the BRST transformations (\ref{BRST2}) reduce to 
\beq
  sX^M = \psi^M,\,\,\,\,\,\,\,\,\,\,
  s\psi^M = 0,\,\,\,\,\,\,\,\,\,\,
  s\bar{\psi}^M =  \dot{X}^M - J^M_{\,\,\,N} \d_1 X^N + \Gamma^M_{\,\,\,NL}
   \bar{\psi}^N \psi^L,
\eeq{newbrstai}
 where $J^M_{\,\,\,N}$ is the complex structure on $M_6$ such that $\omega_{NM} =
   - g_{NL}J^{L}_{\,\,\,M}$. 
Let us introduce the complex coordinates $I=(i, \bar{i})$ with respect to $J$ and redefine our
 fields as follows
\beq
 \psi^i = i\alpha \chi^i,\,\,\,\,\,\,\,\,\,\,
 \psi^{\bar{i}} = i\tilde{\alpha} \chi^{\bar{i}},\,\,\,\,\,\,\,\,\,\,
 \bar{\psi}^i = - \frac{1}{\tilde{\alpha}} \psi^i_{\bar{z}},\,\,\,\,\,\,\,\,\,\,
 \bar{\psi}^{\bar{i}} = - \frac{1}{\alpha} \psi^{\bar{i}}_{z},
\eeq{newfieldsde}
 where $\alpha$ and $\tilde{\alpha}$ are some non-zero constants. In the complex coordinates 
 and new fields the action (\ref{fullaction1}) becomes
\beq
     S_{GF}= k \int d^2\sigma \left ( \frac{1}{2} 
     \d_z X^N g_{NM} \d_{\bar{z}} X^M  
     + i\psi^i_{\bar{z}} g_{i\bar{j}} \nabla_z \chi^{\bar{j}} + i \psi^{\bar{i}}_{z} g_{\bar{i}j} \nabla_{\bar{z}}
  \chi^j - {\cal R}_{i\bar{k}l{\bar{s}}} \psi^i_{\bar{z}}
      \psi^{\bar{k}}_{z} \chi^l \chi^{\bar{s}} \right ),
   \eeq{fullaction2}
 where we introduced $\d_z=\d_0 +i \d_1$, $\nabla_z = \nabla_0 + i \nabla_1$ and their complex conjugates.  
 In the new notation the BRST transformations become
$$ s X^i = i \alpha \chi^i ,\,\,\,\,\,\,\,\,\,\,\,
 sX^{\bar{i}} = i\tilde{\alpha} \chi^{\bar{i}},\,\,\,\,\,\,\,\,\,\,\,
s\chi^i=s \chi^{\bar{i}} =0,\,\,\,\,\,\,\,\,\,\,\,$$
\beq
s \psi^{i}_{\bar{z}} = - \tilde{\alpha} \d_{\bar{z}} X^i - i\alpha \Gamma^i_{\,\,\,nl}\chi^n \psi^l_{\bar{z}},\,\,\,\,\,\,\,\,\,\,\,
 s \psi^{\bar{i}}_z = -\alpha \d_z X^{\bar{i}} - i\tilde{\alpha} \Gamma^{\bar{i}}_{\,\,\,\bar{n}\bar{l}}
  \chi^{\bar{n}} \psi_z^{\bar{l}}.
\eeq{newBRSTW}
 Indeed the action (\ref{fullaction2}) and the transformations (\ref{newBRSTW})
  are exactly the same as the topological A-model in  
  \cite{Witten:1991zz}. 

Next consider the membranes which are localized on $S^1$,
 i.e. $X^7=const.$, $\psi^7=0$ and $\bar{\psi}^7=0$.  In this case the action (\ref{fullaction}) is reduced to
$$     S_{GF}=\int d^3\sigma \left ( \frac{1}{2} \dot{X}^N g_{NM}\dot{X}^M+
  \frac{1}{2}\det (\d_a X^N g_{NM} \d_b X^M)
     - \bar{\psi}^N g_{NM} \nabla_0 \psi^M \right .$$
     \beq
 \left .    - (Re\,\Omega)_{MNL} \bar{\psi}^M \nabla_a \psi^N
      \d_b X^L \epsilon^{ab} + \frac{1}{4} {\cal R}_{MNLS} \bar{\psi}^M
      \psi^S \bar{\psi}^N \psi^L \right )
   \eeq{fullaction3}  
    where we have used (\ref{instajdpel}) and (\ref{metrifcnso}). The BRST transformation 
     becomes
     \beq
    sX^M = \psi^M,\,\,\,\,\,\,\,\,\,\,
  s\psi^M =0,\,\,\,\,\,\,\,\,\,\,
  s\bar{\psi}^M =  \dot{X}^M + (Re\,\Omega)^M_{\,\,\,NL} \d_1 X^N \d_2 X^L 
   + \Gamma^M_{\,\,\,NL}
   \bar{\psi}^N \psi^L  .
\eeq{BRST6dimmem}
 The action (\ref{fullaction3}) is $s$-exact modulo a topological term, i.e.
 \beq
  S_{GF} = \frac{1}{6} \int X^*(Re\,\Omega) + \frac{1}{2}
   \int d^3\sigma\,\, s \left ( \bar{\psi}^M (g_{MN} \dot{X}^N 
      + (Re\,\Omega)_{MNL} \d_1 X^N \d_2 X^L)\right ).
\eeq{local3dilqp-1}
 This membrane theory is localized on the configurations 
 \beq
 \dot{X}^M 
      + (Re\,\Omega)^{M}_{\,\,\,NL} \d_1 X^N \d_2 X^L =0,
 \eeq{confkspq123}
 which are special Lagrangian submanifolds with phase zero (i.e., calibrated by $Re\,\Omega$).

 If on $M_6 \times S^1$ we choose the different $G_2$-structure (\ref{familyG2str}) then the 
 membrane theory on $M_6$ would be a bit different: in all equations
 (\ref{fullaction3})-(\ref{confkspq123}) $Re\,\Omega$ should be replaced by $Re\,(e^{i\theta}\Omega)$. 
 Now the membrane theory is localized on special Lagrangian manifolds with phase $\theta$. 

 To conclude this Section we would like to make a comment on Mathai-Quillen formalism. 
 Indeed the action (\ref{fullaction}) could be constructed within this formalism 
 if we would choose the condition  (\ref{definosnal}) as the appropriate zero section.  

\section{Hamiltonian treatment}
\label{ham}

In this Section we sketch the Hamiltonian treatment of the model (\ref{fullaction}). Indeed 
the Hamiltonian formalism is useful for the geometrical interpretation of BRST transformations
 (\ref{BRST2}) and (\ref{BRST2extra}).

Starting from the gauge fixed action (\ref{fullaction}) we define the momenta
\beq
 p_\mu = g_{\mu\nu} \dot{X}^\nu - \Gamma_{\lambda\mu\nu} \bar{\psi}^\lambda \psi^\nu,
 \,\,\,\,\,\,\,\,\,\,\,\,\,\,\,\,\,
 p_{\psi\mu} =  g_{\mu\nu} \bar{\psi}^\nu,
\eeq{definmomenta}
 where we defined the odd momenta $p_\psi$ using the right derivative of $S_{GF}$ 
 with respect to $\d_0 \psi^\mu$.  The canonical commutation relations are 
 \beq
 \{ X^\mu (\sigma), p_\nu(\sigma') \} = \delta^\mu_{\,\,\,\nu} \delta^2(\sigma -\sigma'),
 \,\,\,\,\,\,\,\,\,\,\,\,\,\,\,\,\,\,\,
 \{\psi^\mu(\sigma), \bar{\psi}_\nu(\sigma')\}_+ = \delta^\mu_{\,\,\,\nu} \delta^2 (\sigma-\sigma')
 \eeq{cancomrelspa}
 with other being trivial and $\{\,\,,\,\,\}_+$ denotes the odd Poisson bracket.
  The Hamiltonian $H_{GF}$ corresponding to $S_{GF}$ is obtained by Legendre 
   transform and can be written as follows
$$ H_{GF} = \int d^2\sigma\,\, \left ( \frac{1}{2} (p_\mu + \Gamma_{\lambda\mu\rho}\bar{\psi}^\lambda \psi^\rho)
 g^{\mu\nu} (p_\nu + \Gamma_{\sigma\nu\tau} \bar{\psi}^\sigma \psi^\tau) - \frac{1}{2} \det (\d_a X^\mu g_{\mu\nu}
 \d_b X^\nu) \right .$$ 
\beq
\left . +\Phi_{\mu\nu\rho} \bar{\psi}^\mu \nabla_a \psi^\nu \d_b X^\rho \epsilon^{ab} -
 \frac{1}{4} {\cal R}_{\mu\sigma\lambda\rho} 
 \bar{\psi}^\mu \bar{\psi}^\sigma \psi^\lambda \psi^\rho \right ).
\eeq{fullhamiltonian}
 In the phase space the generator of BRST transformations (\ref{BRST2}) is 
\beq
 Q = \int d^2\sigma\,\,\psi^\mu (p_\mu + \Phi_{\mu\nu\rho} \d_1 X^\nu \d_2 X^\rho ),
\eeq{definQ}
 where one should be careful in working with the contravariant and covariant versions of 
  $\bar{\psi}$.
The anti-BRST transformations (\ref{BRST2extra}) are generated by
\beq
 \bar{Q} = \int d^2\sigma\,\, \bar{\psi}_\tau g^{\tau\mu}
  (p_\mu - \Phi_{\mu\nu\rho} \d_1 X^\nu \d_2 X^\rho 
 + \Gamma_{\sigma\mu\rho} \bar{\psi}^\sigma \psi^\rho ).
\eeq{deftibarsQ}
 Indeed $\bar{Q}$ is minus adjoint operator of $Q$. This can be 
  shown directly if $Q$ is understood  as an operator on the states $K_{\mu_1...\mu_r}
   \psi^{\mu_1}...\psi^{\mu_r}|0\rangle $ where $K$ is an r-form on $M$
and $\bar\psi^\mu|0\rangle=0$. The inner product 
    of two forms is defined as usual, $\int K' \wedge * K$,
with $*K= (*K)_{\mu_1\ldots\mu_{d-r}}\bar\psi^{\nu_1}\ldots\bar\psi^{\nu_r}
\epsilon^{\mu_1\ldots\mu_{d-r}}_{\,\,\,\,\,\,\,\,\,\,\,\,\,\,\,\,\,\,\,\,\,\, \nu_1\ldots\nu_r}$, where $d=7$ is the
space-time dimension.
The operator $Q$ acts on the differential forms as a de Rham 
differential and $\bar{Q}$ as minus its adjoint. 
This explains the choice of the bilinear fermionic term in the gauge function 
(\ref{gaugefunct}). In fact,
analogously to the discussion in \cite{Baulieu:1989rs} one can check that this
the only choice for which $\bar Q = -Q^\dagger$.

  The Hamiltonian (\ref{fullhamiltonian}) $H_{GF}$, the BRST generator $Q$ 
  and the anti-BRST generator $\bar{Q}$ satisfy the following on--shell 
  relations
  \beq
   H_{GF} = \frac{1}{2} \{ Q, \bar{Q}\}_+,\,\,\,\,\,\,\,\,\,\,\,\,\,\,
   Q^2=0,\,\,\,\,\,\,\,\,\,\,\,\,\,\,
   \bar{Q}^2=0.
  \eeq{definHamds}
 As result of this $H_{GF}$ is BRST and anti-BRST invariant, 
 i.e. $\{H_{GF}, Q\}=\{H_{GF}, \bar{Q}\}=0$.

In principle we could proceed with the construction of $S_{GF}$ via 
Hamiltonian formalism.
The model is described by the following first class constraints
 \beq
 J_\mu = p_\mu + \Phi_{\mu\nu\rho} \d_1 X^\nu \d_2 X^\rho,
 \eeq{firstclassconstao}
which has been discussed in \cite{Bonelli:2005ti}.  
Since these constraints satisfy the following Poisson brackets 
   \beq
   \{ J_\mu(\sigma), J_\nu (\sigma') \} = 0,
   \eeq{algebatrdoa}
introducing the ghosts $\psi$ 
one can construct the BRST charge in the minimal sector 
\cite{ennoteca} (\ref{definQ}). There should exist a non--minimal
extension and a suitable gauge--fixing such that after integrating out
the non--minimal sector one recovers the gauge--fixed BRST
structure described above. 
 Some of the aspects of the Hamiltonian analysis of this and related systems 
 has been discussed in \cite{Bonelli:2005ti}. Indeed the correct reduction 
 of this model on $M_6 \times S^1$ works at the level  constraints  
 (\ref{firstclassconstao}) as well. 
 
 Moreover the Hamiltonian point of view suggests that if one wishes to 
 include the flux
 $4$-form $H$ into consideration then the topological model is defined as
 \beq
  S_{top} = \int\limits_{\Sigma_3} X^*(\Phi) - \int\limits_{\Sigma_4} X^*(H),
 \eeq{definfluxtop}
 such that $\Phi$ is a three--form associated with a $G_2$-structure, 
 $d\Phi=H$ and $\d\Sigma_4=\Sigma_3$ \footnote{The action 
 (\ref{definfluxtop}) is invariant due to the identity
 $$\delta\int\limits_{\Sigma_n}X^*(\Phi)= \int\limits_{\partial\Sigma_n}X^*(i_{\delta X}
\Phi) + \int\limits_{\Sigma_n}X^*(i_{\delta X}d\Phi)$$.}.
  The dimensional reduction on $M_7=M_6\times S^1$ of this model
 should give a topological sigma model for generalised complex geometries. 
  We hope to discuss this extension of our model elsewhere.

\section{Observables and moduli spaces}
\label{observables}

In this Section we collect some generalities on the topological 
membrane theory constructed in Section \ref{gfaction}.
 We start by discussing the observables in the theory. 
For a nontrivial element $[K] \in H^q(M)$ we 
can formally define the following cocycles
\ber
&&{\cal C}_3^{q-3} = \frac{1}{6} K_{\mu_1...\mu_q} dX^{\mu_1} \wedge dX^{\mu_2} \wedge dX^{\mu_3}
  \psi^{\mu_4}...\psi^{\mu_q} , \nonumber\\
 && {\cal C}_2^{q-2} = \frac{1}{2} K_{\mu_1...\mu_q} dX^{\mu_1} \wedge dX^{\mu_2} \psi^{\mu_3}
  \psi^{\mu_4}...\psi^{\mu_q} ,\nonumber\\
&&{\cal C}_1^{q-1} = K_{\mu_1...\mu_q} dX^{\mu_1} \psi^{\mu_2} \psi^{\mu_3}
  \psi^{\mu_4}...\psi^{\mu_q} ,\nonumber\\
&&{\cal C}_0^{q} = K_{\mu_1...\mu_q} \psi^{\mu_1} \psi^{\mu_2} \psi^{\mu_3}
  \psi^{\mu_4}...\psi^{\mu_q}, 
\eer{closedforma}
 where in ${\cal C}_i^{q-i}$ the upper index stands for the ghost number 
 and the lower index for the degree of the differential form on $\Sigma_3$. 
 Using the transformations (\ref{BRST1}) we can derive 
 the decent equations for ${\cal C}_i^{q-i}$
   \beq
   s {\cal C}_3^{q-3} = \frac{1}{q-2} d {\cal C}_2^{q-2},\,\,\,\,\,\,\,\,\,\,\,\,\,
    s {\cal C}_2^{q-2} = \frac{1}{q-1} d {\cal C}_1^{q-1},\,\,\,\,\,\,\,\,\,\,\,\,\,
 s {\cal C}_1^{q-1} = \frac{1}{q} d {\cal C}_0^{q},\,\,\,\,\,\,\,\,\,\,\,\,\,
 s {\cal C}_0^{q} = 0.
    \eeq{deceqsk}
   Thus ${\cal C}_0^q$ are BRST-invariant local observables labeled by 
   the elements of the de Rham complex $H^\bullet (M)$. 
   From ${\cal C}_i^{q-i}$ with $ i >0$ we can construct 
     BRST-invariant non-local observables as integrals
     \beq
     \int\limits_{c_i} {\cal C}_i^{q-i} \, ,
     \eeq{integracycle}
      where $c_i$ is $i$-cycle on $\Sigma_3$. 
However not all observables have non-vanishing 
correlators in the theory. 
To study this we need to address the ghost number anomaly.
The action (\ref{fullaction}) has at the classical level a ghost 
number conservation law, with $\psi$ having ghost number $1$, $\bar{\psi}$ 
having ghost number $-1$ and $X$ having ghost number $0$. The BRST transformation  $s$ (\ref{BRST2}) changes the ghost number by $1$. 
Notice that all the observables, but ${\cal C}_i^0$ with $i=1,2,3$, defined in (\ref{closedforma})
have a non--vanishing ghost number. Thus, in order to have non--vanishing
correlators there should be a compensating ghost number anomaly.
The linearized equations for the fermionic fluctuations
around the instanton background are
\ber
&&D\psi^\mu= \nabla_0 \psi^\mu + \Phi^\mu_{\,\,\,\nu\rho} \epsilon^{ab} \nabla_a \psi^\nu \d_b X^\rho =0,\label{zeromod22}\\
 &&D^\dagger \bar{\psi}^\mu =
 \nabla_0 \bar{\psi}^\mu - \Phi^\mu_{\,\,\,\nu\rho} \epsilon^{ab} \nabla_a \bar{\psi}^\nu \d_b X^\rho =0.
  \label{zeromod11}
\eer{zeromodelala}
The equation (\ref{zeromod22}) is the first order variation of the associative map 
(\ref{definosnal}). As such, $\psi$   
can be interpreted as a section of the tangent bundle to 
the moduli space ${\cal M}$ of associative maps.  
Indeed the operator  $D^\dagger$ is the adjoint of $D$, 
and thus the ghost number anomaly is given by the index $ind(D)$.  
Since our theory lives in three dimensions 
 $ind(D)$ vanishes by index theorem. 
Thus the correlators of the operators in 
(\ref{closedforma}) are vanishing, except $C^0_i$ with $i=1,2,3$,
where $C^0_3$ corresponds to our classical action (\ref{topaction}) and
its variations in $H^3(M)$.
The  non--trivial topological observable is the partition
function, which computes the Euler characteristic of the moduli
space of associative maps. In fact,
the zero modes of (\ref{zeromod22}) and (\ref{zeromodelala}) can be soaked up by 
the ${\cal R} \psi^2 \bar{\psi}^2$ term
of the action (\ref{fullaction}). 
The situation is very similar to supersymmetric quantum 
mechanics (for a review, see \cite{Birmingham:1991ty}).

The above discussion was rather formal and 
we did not analyse the moduli space
of associative maps. Not so much is known about 
these moduli spaces, see
for example the work by McLean \cite{McLean} and the recent works 
\cite{akbulut1, akbulut2}.  
Actually our Dirac operator $D$ 
can be mapped to McLean operator since both deal with 
the deformations of associative 
manifolds. Indeed, instead of (\ref{definosnal}) we could have chosen
a different parametrization for the associative maps corresponding
to the static gauge
\ber
&& X^\alpha = \sigma^\alpha \, , \\
&& \dot X^i \pm \Phi^i_{\,\,\,\,aj} \epsilon^{ab} \partial_b X^j = 0 \, ,
\eer{static}
where $\alpha=0,1,2$ and $i=3,\ldots,6$.
The BRST variation of (\ref{static}) gives the McLean twisted Dirac operator.
In the gauge (\ref{static}) our model reproduces the results of 
the supermembrane theory in \cite{Beasley:2003fx} and also \cite{Anguelova:2005cv}.
      
Although the membrane theory does not display a ghost number anomaly,
it suffers a global anomaly under parity transformations \cite{Atiyah:2001qf}.
This has a counterpart in the A--model upon $S^1$ compactification.
For membranes wrapping the $S^1$ with a given winding number $k$
the spectral flow of the operator
(\ref{zeromod22}) should match the index of the $\nabla_z$ operator
in (\ref{fullaction2}) by arguments similar to those in 
\cite{Alvarez-Gaume:1984nf}. This should allow to recover the ghost
number anomaly of the A model and the corresponding non--trivial
correlators coming from (\ref{closedforma}).

\section{Vector cross products and TFTs}
\label{vcp}

In Sections \ref{membrane} and \ref{gfaction} we discussed the membrane instantons 
 on $G_2$-manifolds and constructed topological membrane theory which localizes on 
  these instantons. The whole construction is very similar to A-model (topological sigma model)
   \cite{Witten:1988xj, Baulieu:1989rs}. Indeed there is whole set of topological $p$-brane 
   models which follows the same pattern.  These models are based on the geometrical notion of vector cross product structure. In what follows we sketch the main steps of the construction of topological 
    theories based both on real and complex vector cross products. 

 We start with the case of a real cross vector product (see Appendix for the definition and properties).  
 Consider the Nambu-Goto $p$-brane theory on the manifold $M$ with Riemannian metric $g$ 
 \beq
 S = \int d^{p+1}\sigma\,\,\sqrt{\det(\d_\alpha X^\mu g_{\mu\nu} \d_\beta X^\nu)},
 \eeq{Naameppp}
  where $\alpha=0,1,...,p$.
  In analogy with the membrane case there is a gauge in which the $p$-brane action has
  the following form
   \beq
    S = \frac{1}{2} \int d^{p+1}\sigma\,\,\left ( \dot{X}^\mu g_{\mu\nu} \dot{X}^\nu +
       \det (\d_a X^\mu g_{\mu\nu} \d_b X^\nu) \right ),
\eeq{gehak12903}
  with $\dot{X}^\mu =\d_0 X^\mu$ and $a,b=1,...,p$.  Assuming that there is 
  a $(p+1)$-form on $M$ we can write down the bound
  \beq
\int d^{p+1}\sigma\,\, \left ( \dot{X}^\mu \pm \phi^\mu_{\,\,\,\nu_1...\nu_p} \d_1 X^{\nu_1}...
 \d_p X^{\nu_p} \right ) g_{\mu\lambda}
 \left ( \dot{X}^\lambda \pm \phi^\lambda_{\,\,\,\sigma_1...\sigma_p}\d_1 X^{\sigma_1}...
  \d_p X^{\sigma_p}
  \right ) \geq 0 .
\eeq{boudnao20AAA}      
  If $\phi$ and $g$ correspond to a vector cross product structure on $M$
  then the bound (\ref{boudnao20AAA}) can be rewritten as follows
\beq
 \frac{1}{2} \int d^{p+1}\sigma\,\,\left ( \dot{X}^\mu g_{\mu\nu} \dot{X}^\nu +
       \det (\d_a X^\mu g_{\mu\nu} \d_b X^\nu) \right ) \geq \mp \frac{1}{(p+1)!} \int X^*(\phi).
\eeq{membrabound1AAA}
 Moreover if $d\phi=0$ the right-hand side is a topological term. The bound (\ref{membrabound1AAA})
  is saturated if 
 \beq
 \dot{X}^\mu \pm \phi^\mu_{\,\,\,\nu_1 ... \nu_p} \d_1 X^{\nu_1}... \d_p X^{\nu_p} =0,
\eeq{generinstanton} 
 which we call $p$-brane instanton. Geometrically it corresponds to a submanifold of $M$
  calibrated by $\phi$ \cite{complex}.
 
 Following the considerations from Section \ref{membrane} we consider the topological 
  $p$-brane theory
 \beq
  S_{top} = - \frac{1}{(p+1)!} \int  X^*(\phi),
 \eeq{topactiongen}
  where $\phi$ is a closed $(p+1)$-form corresponding to a cross vector product on $M$. 
 The action (\ref{topactiongen})  is invariant under the gauge symmetry $\delta X^\mu =\epsilon^\mu$. 
  The corresponding BRST transformations are defined as in (\ref{BRST1}). 
   Choosing the gauge function as
   \beq
    {\cal F}^\mu = \dot{X}^\mu + \phi^\mu_{\,\,\,\nu_1...\nu_p} \d_1 X^{\nu_1}...\d_p X^{\nu_p}
     + \frac{1}{2} \Gamma^\mu_{\,\,\,\sigma\rho} \bar{\psi}^\sigma \psi^\rho
   \eeq{gaugefixfubcgen} 
  we define the gauge fixed action as follows
$$ S_{GF} =-\frac{1}{(p+1)!} \int X^*(\phi) + 
 \int d^{p+1}\sigma\,\, s \left ( \bar{\psi}^\mu (g_{\mu\nu} \dot{X}^\nu 
      + \phi_{\mu{\nu_1}...{\nu_p}} \d_1 X^{\nu_1} ...\d_p X^{\nu_p} \right . $$
\beq     
\left .       + \frac{1}{2} \Gamma_{\mu\sigma\rho}
       \bar{\psi}^\sigma \psi^\rho - \frac{1}{2} g_{\mu\nu} b^\nu ) \right ) .
  \eeq{gaugeficm119}
  Eliminating $b$ by its algebraic equation we arrive to the following gauge fixed action
\ber
&& S_{GF}=  \int d^{p+1}\sigma \left( \frac{1}{2} \dot{X}^\mu g_{\mu\nu} \dot{X}^\nu
  + \frac{1}{2}\det (\d_a X^\mu g_{\mu\nu} \d_b X^\nu)
     - \bar{\psi}^\mu g_{\mu\nu} \nabla_0 \psi^\nu  \right. \nonumber\\
&& \left.  + \frac{1}{4} {\cal R}_{\mu\sigma\lambda\rho} \bar{\psi}^\mu
      \psi^\rho \bar{\psi}^\sigma \psi^\lambda  
      -\frac{1}{(p-1)!}\phi_{\mu{\nu_1}{\nu_2}...{\nu_p}} \bar{\psi}^\mu \nabla_{a_1} \psi^{\nu_1}
      \d_{a_2} X^{\nu_2} ...\d_{a_p} X^{\nu_p}
       \epsilon^{a_1 a_2 ... a_p} \right. \nonumber\\
&& \left . -\frac{1}{(p-2)!}\nabla_\lambda \phi_{\mu{\nu_1}{\nu_2}...{\nu_p}} \bar{\psi}^\mu 
      \psi^\lambda \d_{a_1} X^{\nu_1}...\d_{a_p} X^{\nu_p}
       \epsilon^{a_1 a_2 ... a_p} \right),
  \eer{gjdkwp293}
  where $\nabla_\alpha \psi$ is defined in (\ref{nbladef}).
The action (\ref{gjdkwp293}) is invariant under the following BRST transformations
\beq
sX^\mu = \psi^\mu,\,\,\,\,\,\,\,\,\,\,
  s\psi^\mu =0,\,\,\,\,\,\,\,\,\,\,
  s\bar{\psi}^\mu =  \dot{X}^\mu + \phi^\mu_{\,\,\,\nu_1 ... \nu_p} \d_1 X^{\nu_1}... \d_p X^{\nu_p} + \Gamma^\mu_{\,\,\,\sigma\rho}
   \bar{\psi}^\sigma \psi^\rho ,
\eeq{BRSTgenrpq}
 which are nilpotent on-shell. The action (\ref{gjdkwp293}) can be rewritten as
 \beq
  S_{GF} =  - \frac{1}{(p+1)!} \int  X^*(\phi) + \frac{1}{2} \int d^{p+1}\sigma\,\, s\left (
  \bar{\psi}^\mu (g_{\mu\nu} \dot{X}^\nu 
      + \phi_{\mu{\nu_1}...{\nu_p}} \d_1 X^{\nu_1} ...\d_p X^{\nu_p} ) \right )
 \eeq{fghs03wr888}
 and therefore the model is localized on the $p$-brane instantons
  (\ref{generinstanton}).  Due to standard arguments the theory does not depend
   on the way we describe $p$-brane instantons. Namely any change in the 
    distinguished direction $\alpha=0$ in the path integral will contribute a BRST--exact term 
     and thus is irrelevant.  
 
 An interesting question is: how generic is our construction? Actually, all real vector cross products
   have been classified   by Brown and Gray \cite{brown} (see the list in the Appendix B). 
    There are four different cases for which the cross product exists. The first case corresponds 
     to $\phi$ being the volume form on $M$. In this case the TFT we constructed corresponds
      to $p$-branes embedded into a $p+1$ dimensional space $M$. Some of the aspects 
       of this TFT has been discussed in \cite{Bengtsson:2000xa}. The second case corresponds
        to a symplectic manifold with $\phi$ being a closed non-degenerate $2$-form.
         The corresponding TFT is just topological sigma model (A-model)
          \cite{Witten:1988xj}.  The remaining two vector cross product structures are the exceptional 
 cases.  The first corresponds to seven dimensional manifolds with 
   $G_2$-structure and $\phi$ is the three form $\Phi$. This is the theory we constructed 
    in Section \ref{gfaction}.  The second exceptional case corresponds to eight dimensional 
     manifolds with $Spin(7)$-structure where $\phi$ the associated $4$-form $\Psi$ (the Cayley form).    
      In this case our model describes $3$-branes in a $Spin(7)$-manifold. This  is
      presumably the  microscopic description of the
        recently proposed  topological F-theory \cite{Anguelova:2004wy}. 
         Therefore we refer to this theory as topological F-theory on $Spin(7)$-manifolds. 
        This TFT is localized 
         on Cayley 4-manifolds (i.e, those calibrated by $\Psi$).
    It is not hard to repeat for the topological F-theory the analysis which we have done 
     in Sections \ref{gfaction}-\ref{observables} for topological M-theory. In particular 
      one can consider the reduction of F-theory on 
 $$   CY_3 \times T^2\,\,\,\,\Longrightarrow\,\,\,\,CY_3 \times S^1\,\,\,\,\Longrightarrow\,\,\,\,
  CY_3,$$   
    where $CY_3 \times T^2$ is a $Spin(7)$-manifold. This reduction will produce the whole 
     Zoo of TFTs on $CY_3$ which were discussed briefly at the Hamiltonian level 
      in \cite{Bonelli:2005ti}. One can perform the reduction also at the level of the gauge 
       fixed action in a similar way as in Section \ref{gfaction}.

   So far we have considered the real cross vector product structures. On Hermitian manifolds
    $(g, J, M)$  one can introduce the complex version \cite{complex} 
     of cross vector products (see the Appendix B
     for the definition). In this case the complex vector product\footnote{If $M$ is a K\"ahler
      manifold with complex vector cross product then $M$ is either Calabi-Yau with a holomorphic 
       volume form or hyperk\"ahler with holomorphic symplectic form.}
      is given by a holomorphic
      $p$-form which is either a holomorphic volume form or a holomorphic symplectic form on $M$.
      Indeed it is straightforward to generalize our construction
      to the topological action
      \beq
       S_{top} = -\frac{1}{(p+1)!} \int  X^*(Re(e^{i\theta}  \Omega)),
      \eeq{comvectpr} 
      where $\Omega$ is a closed form corresponding to a complex vector cross product on 
       $(g, J, M)$. 
        The construction of the gauge fixed action goes along the lines we have presented 
        above and thus we give only the final result of the construction. The gauge fixed action is 
\ber
&& S_{GF}=  \int d^{p+1}\sigma \left ( \frac{1}{2} \dot{X}^\mu g_{\mu\nu} \dot{X}^\nu
  + \frac{1}{2}\det (\d_a X^\mu g_{\mu\nu} \d_b X^\nu)
     - \bar{\psi}^\mu g_{\mu\nu} \nabla_0 \psi^\nu  \right . \nonumber\\
&& \left .  + \frac{1}{4} {\cal R}_{\mu\sigma\lambda\rho} \bar{\psi}^\mu
      \psi^\rho \bar{\psi}^\sigma \psi^\lambda -\frac{1}{(p-1)!}
(Re (e^{i\theta}\Omega))_{\mu{\nu_1}{\nu_2}...{\nu_p}} \bar{\psi}^\mu \nabla_{a_1} \psi^{\nu_1}
      \d_{a_2} X^{\nu_2} ...\d_{a_p} X^{\nu_p}
       \epsilon^{a_1 a_2 ... a_p} \right. \nonumber\\
&& \left.  -\frac{1}{(p-2)!}
\nabla_\lambda(Re (e^{i\theta} \Omega))_{\mu{\nu_1}{\nu_2}...{\nu_p}} \bar{\psi}^\mu \psi^\lambda
\d_{a_1} X^{\nu_1}
      \d_{a_2} X^{\nu_2} ...\d_{a_p} X^{\nu_p}
       \epsilon^{a_1 a_2 ... a_p}
\right ),
  \eer{gjdkwp293complex}
     which is invariant under the following BRST transformation
    \beq
sX^\mu = \psi^\mu,\,\,\,\,\,\,\,\,\,\,
  s\psi^\mu =0,\,\,\,\,\,\,\,\,\,\,
  s\bar{\psi}^\mu =  \dot{X}^\mu + (Re (e^{i\theta}\Omega))^\mu_{\,\,\,\nu_1 ... \nu_p} \d_1 X^{\nu_1}... \d_p X^{\nu_p} + \Gamma^\mu_{\,\,\,\sigma\rho}
   \bar{\psi}^\sigma \psi^\rho .
\eeq{BRSTgenrpqcomplex}
 In this construction it is essential that the metric $g$ is Hermitian and $\Omega$ is either a
  holomorphic symplectic form or a holomorphic volume form. The present model is localized on 
  submanifolds calibrated by $Re(e^{i\theta}\Omega)$.

\section{Conclusions}
\label{end}

In this work we have constructed the gauge fixed action for the topological membrane 
on $G_2$-manifolds. The bosonic part of the action is the standard membrane action 
in a particular gauge. This TFT is localized on associative maps
and its partition function computes the Euler characteristic
of the corresponding moduli space\footnote{For example,  
 the contribution of the constant maps to the free energy  
  of both our model and the one presented in \cite{Anguelova:2005cv} does not give 
 the volume of the $G_2$-manifold due to presence of fermionic 
  zero modes.}.

Indeed our model plays the analogous r\^ole for topological M theory as the topological
sigma model for the topological string.
Therefore in order to complete the program of giving a microscopic
description of topological M theory in terms of membranes,
a crucial issue is the coupling with three--dimensional topological gravity.
In analogy with the topological A string,
the contribution of the constant maps to the partition function
of this complete membrane model should give the volume of the target $G_2$ manifold,
and hence the Hitchin functional considered in \cite{Dijkgraaf:2004te}. 

The coupling of our model to 3D gravity requires a covariant world--volume
gauge--fixed description. The relevant three--dimensional gauge theory
should be a $BF$ theory with $SU(2)$ gauge group\footnote{The possible
relevance of $BF$ theory in the context of topological M theory was
put forward in \cite{Dijkgraaf:2004te,Baulieu:2004pv}.}.
This coupling will contribute to the three--dimensional parity anomaly
in such a way that the complete model on $S^1$ at fixed winding 
will match the ghost anomaly of the topological string and hopefully
reproduce the non--trivial correlators of the latter theory. 

We also generalized our approach to topological $p$-brane theories  
corresponding to real and complex vector cross product structures on $M$.  
In particular, there is a well--defined topological $3$--brane theory on $Spin(7)$--manifolds,
which is possibly relevant for topological F theory and whose quantum mechanical
properties deserve further study.  
 Notice that we expect this theory to display
a non-vanishing ghost anomaly, thus completing the analogous of the dimensional
ladder of \cite{Alvarez-Gaume:1984nf,Forte:1986em}.

\noindent{\bf Acknowledgement}:  We are grateful to Lilia Anguelova, 
 Glenn Barnich,
 Matthias Blau, Paul de Medeiros,
 Sergei Gukov and Nigel Hitchin and Annamaria Sinkovics for discussions. 
 The research of G.B. is supported by the Marie Curie European 
 Reintegration Grant MERG-CT-2004-516466,
 the European Commission RTN Program MRTN-CT-2004-005104 and by MIUR.
 M.Z. thanks SISSA (Trieste) and Simons workshop (Stony Brook)
   where part of this work was carried out. 
 The research of M.Z. was supported by EU-grant
 MEIF-CT-2004-500267 and VR-grant 621-2004-3177.

\appendix 
 
\Section{$G_2$-manifolds} 
\label{G2} 

In this Appendix we collect the relevant information about seven dimensional manifolds
 with $G_2$-structure. For further details the reader may consult \cite{joyce}.
 
 Let $e_1,e_2,...,e_7$ denote the standard basis of $\mathbb{R}^7$ and let $e^1,e^2,....,e^7$
  denote the corresponding dual basis. Define an element in $\Lambda^3((\mathbb{R}^7)^*)$
  $$\Phi_0 = e^{123} + e^{145} + e^{167} + e^{246} -e^{257} - e^{347} - e^{356},$$
  where $e^{ijk}\equiv \frac{1}{3!}e^i\wedge e^j \wedge e^k$. The group $G_2$ is defined as follows
  $$ G_2 = \{ g \in GL(7, \mathbb{R}), g^*(\Phi_0)=\Phi_0 \},$$
  i.e. $G_2$ is the stabilizer subgroup of $\Phi_0$ in $GL(7,\mathbb{R})$.
  
  A smooth seven dimensional manifold $M$ has $G_2$-structure if its tangent frame 
   bundle reduces to a $G_2$ bundle. Equivalently, $M$ has a $G_2$-structure
    if there is a three form $\Phi \in \Omega^3(M)$ such that at each point $x\in M$
     the pair $(T_xM, \Phi_x)$ is isomorphic to $(T_0 \mathbb{R}^7, \Phi_0)$.
     
     A manifold with $G_2$-structure $(M, \Phi)$ is called $G_2$-manifold if the holonomy 
      group of the Levi-Civita connection of the metric $g$ lies inside of $G_2$. Equivalently 
      $(M,\Phi)$ is a $G_2$-manifold if $d\Phi = d *\Phi =0$.
      
     The crucial property of $\Phi$ and $g$ on manifolds with $G_2$--structure 
     we use in the calculation is the following one
     $$ \Phi_{\mu\nu\rho} u^\nu v^\rho \Phi^\mu_{\,\,\,\lambda\sigma} u^\lambda v^\sigma
     = \det \left ( \begin{array}{ll}
                                                  u^\mu g_{\mu\nu} u^\nu  & u^\mu g_{\mu\nu} v^\nu\\
                                                  v^\mu g_{\mu\nu} u^\nu & v^\mu g_{\mu\nu} v^\nu
                                             \end{array} \right ).$$
     This corresponds to the property that there is a vector cross product structure on $M$, 
     see next Appendix.

\Section{Vector cross product structure}
\label{VCP}

 In this Appendix we review the real and complex vector cross product structures on smooth manifolds. 

We start from the real version of vector cross product. 
  We all are familiar with the usual vector cross product $\times$ of two vectors in $\mathbb{R}^3$, which 
 satisfies 

$\bullet$ $u \times v$ is bilinear and skew symmetric

$\bullet$ $ u \times v \perp u, v$; so $(u\times v) \cdot v=0$  and $(u\times v)\cdot u=0$

$\bullet$ $(u \times v) \cdot (u \times v) = \det \left ( \begin{array}{ll}
                                                  u \cdot u & u \cdot v\\
                                                  v\cdot u & v \cdot v
                                             \end{array} \right )$\\

 The generalization of vector cross product to a Riemannian manifold
  leads to the following definition by Brown and Gray \cite{brown}

\begin{definition}
 On $d$-dimensional Riemannian manifold $M$ with a metric $g$ an $p$-fold vector cross product
 is a smooth bundle map
$$  \chi : \wedge^p TM \rightarrow TM $$
satisfying
$$g (\chi(v_1,...,v_p), v_i)=0,\,\,\,\,\,\,1 \leq i \leq p$$
$$ g(\chi(v_1,...,v_p), \chi(v_1,...,v_p)) = \| v_1 \wedge ... \wedge v_p \|^2$$
 where $\|...\|$ is the induced metric on $\wedge^p TM$.
\end{definition}
 Equivalently the last property can be rewritten in the following form
$$ g(\chi(v_1,...,v_p), \chi(v_1,...,v_p)) = \det ( g(v_i, v_j))= \| v_1 \wedge ... \wedge v_p \|^2. $$
 The first condition in the above definition is equivalent to the following tensor $\phi$
$$ \phi (v_1,...,v_p, v_{p+1}) = g (\chi(v_1,...,v_p), v_{p+1})$$
 being a skew symmetric tensor of degree $p+1$, i.e. $\phi \in \Omega^{p+1}(M)$. 
 Thus in what follows we consider a $(p+1)$-form $\phi$ which defines the $p$-fold vector 
 cross product.  Alternatively a vector cross product form can be defined via a form $\phi \in \Omega^{p+1}(M)$
  satisfying the following property
  $$ \| i_{e_1 \wedge e_2 \wedge ...\wedge e_{p}} \phi \| =1$$
  for any orthonormal set $e_1, e_2, ..., e_p \in T_x M$ and any $x \in M$.

Cross products on real spaces  were classified by Brown and Gray \cite{brown}. 
 The global vector cross products on manifolds were first studied by Gray \cite{gray}.
 They fall into four categories:

(1) $p=d-1$ and $\phi$ is the volume form of the manifold.

(2) $d$ is even and $p=1$. In this case we have a one-fold cross product $J: TM \rightarrow TM$. 
 Such a map satisfies $J^2=-1$ and is an almost complex structure. The associated $2$-form 
 is the K\"ahler form.

(3) The first of two exceptional cases is a $2$-fold cross product ($p=2$) on a $7$-manifold.
 Such a structure  is called a $G_2$-structure and the associated $3$-form is called 
 a $G_2$-form (that is $\Phi$ in the notation of Appendix A and in the main text).

(4) The second exceptional case is $3$-fold cross product ($p=3$) on an
 $8$-manifold. This is called a $Spin(7)$-structure and the associated 
 $4$-form is called $Spin(7)$-form.

  The complex version of vector cross product has been introduced in \cite{complex}.
   Consider a Hermitian manifold $(g, J, M)$ and define the complex vector cross 
    product as a holomorphic $(p+1)$-form satisfying 
    $$ \| i_{e_1 \wedge e_2 \wedge ...\wedge e_{p}} \phi \| =2^{(p+1)/2}$$
for any orthonormal tangent vectors $e_1, e_2,...,e_p \in T^{1,0}_xM$, for any $x\in M$.
One can show from this definition that $\phi$ can be either a holomorphic symplectic form
 or a holomorphic volume form \cite{complex}.  Thus the examples of manifolds equipped with
  the complex vector cross product structure are hyperk\"ahler and Calabi-Yau manifolds.


\begin{thebibliography}{6666} 

\newcommand{\np}{{\em Nucl.\ Phys.\ }} 
\newcommand{\pr}{{\em Phys.\ Rev.\ }} 
\newcommand{\cmp}{{\em Commun.\ Math.\ Phys.\ }} 
\newcommand{\pl}{{\em Phys.\ Lett.\ }} 
%
\bibitem{akbulut1}
S.~Akbulut and S.~Salur,
"Calibrated Manifolds and Gauge Theory,"
arXiv:math.GT/0402368.
%
\bibitem{akbulut2}
S.~Akbulut and S.~Salur,
"Associative submanifolds of a G2 manifold,"
arXiv:math.GT/0412032.
%
\bibitem{Alvarez-Gaume:1984nf}
  L.~Alvarez-Gaume, S.~Della Pietra and G.~W.~Moore,
  ``Anomalies And Odd Dimensions,''
  Annals Phys.\  {\bf 163} (1985) 288.
\bibitem{Anguelova:2004wy}
  L.~Anguelova, P.~de Medeiros and A.~Sinkovics,
  ``On topological F-theory,''
  JHEP {\bf 0505} (2005) 021
  [arXiv:hep-th/0412120].
%
\bibitem{Anguelova:2005cv}
  L.~Anguelova, P.~de Medeiros and A.~Sinkovics,
  ``Topological membrane theory from Mathai-Quillen formalism,''
  arXiv:hep-th/0507089.
%
\bibitem{Atiyah:2001qf}
  M.~Atiyah and E.~Witten,
  ``M-theory dynamics on a manifold of G(2) holonomy,''
  Adv.\ Theor.\ Math.\ Phys.\  {\bf 6} (2003) 1
  [arXiv:hep-th/0107177].
%
\bibitem{Bao:2005yx}
  L.~Bao, V.~Bengtsson, M.~Cederwall and B.~E.~W.~Nilsson,
  ``Membranes for topological M-theory,''
  arXiv:hep-th/0507077.
%
\bibitem{Baulieu:1989rs}
  L.~Baulieu and I.~M.~Singer,
  ``The Topological Sigma Model,''
  Commun.\ Math.\ Phys.\  {\bf 125} (1989) 227.
%
\bibitem{Baulieu:2004pv}
  L.~Baulieu and A.~Tanzini,
  ``Topological symmetry of forms, N = 1 supersymmetry and S-duality on special
  manifolds,''
  arXiv:hep-th/0412014.

%
\bibitem{Beasley:2003fx}
  C.~Beasley and E.~Witten,
  ``Residues and world-sheet instantons,''
  JHEP {\bf 0310} (2003) 065
  [arXiv:hep-th/0304115].
%
\bibitem{Becker:1995kb}
  K.~Becker, M.~Becker and A.~Strominger,
  ``Five-branes, membranes and nonperturbative string theory,''
  Nucl.\ Phys.\ B {\bf 456} (1995) 130
  [arXiv:hep-th/9507158].
  %
\bibitem{Bengtsson:2000xa}
  I.~Bengtsson, N.~Barros e Sa and M.~Zabzine,
  ``A note on topological brane theories,''
  Phys.\ Rev.\ D {\bf 62} (2000) 066005
  [arXiv:hep-th/0005092].
  %
\bibitem{Birmingham:1991ty}
  D.~Birmingham, M.~Blau, M.~Rakowski and G.~Thompson,
  ``Topological field theory,''
  Phys.\ Rept.\  {\bf 209} (1991) 129.
  %
\bibitem{Bonelli:2005ti}
  G.~Bonelli and M.~Zabzine,
  ``From current algebras for p-branes to topological M-theory,''
  JHEP {\bf 0509} (2005) 015
  [arXiv:hep-th/0507051].
%
\bibitem{brown}
R.~B.~Brown and A.~Gray, 
 ``Vector cross products,''
 Comment.\ Math.\ Helv.\  {\bf 42}  (1967) 222-236.
%
\bibitem{Dijkgraaf:2004te}
  R.~Dijkgraaf, S.~Gukov, A.~Neitzke and C.~Vafa,
  ``Topological M-theory as unification of form theories of gravity,''
  arXiv:hep-th/0411073.
  %
%
\bibitem{Forte:1986em}
  S.~Forte,
  ``Explicit Construction Of Anomalies,''
  Nucl.\ Phys.\ B {\bf 288} (1987) 252.
%
\bibitem{Gerasimov:2004yx}
  A.~A.~Gerasimov and S.~L.~Shatashvili,
  ``Towards integrability of topological strings. I: Three-forms on Calabi-Yau
  manifolds,''
  JHEP {\bf 0411} (2004) 074
  [arXiv:hep-th/0409238].
%
\bibitem{Grassi:2004xr}
  P.~A.~Grassi and P.~Vanhove,
  ``Topological M theory from pure spinor formalism,''
  arXiv:hep-th/0411167.
%
\bibitem{gray}
A.~Gray,
 ``Vector cross products on manifolds,''
 Trans.\ Amer.\ Math.\ Soc.\  {\bf 141}  (1969) 465-504;
 errata {\bf 148} (1970) 625.
%
\bibitem{Harvey:1999as}
  J.~A.~Harvey and G.~W.~Moore,
  ``Superpotentials and membrane instantons,''
  arXiv:hep-th/9907026.
%
\bibitem{calib}
R.~Harvey and H.~B.~Lawson, Jr.,
 "Calibrated geometries," Acta Math. {\bf 148} (1982), 47Ð157.
%
\bibitem{ennoteca}
M.~Henneaux and C.~Teitelboim, {\it Quantization of Gauge Systems},
Princeton, USA: Univ. Pr. (1992) 520 p.
%
\bibitem{joyce}
D.~D.~Joyce,
"Compact Manifolds with Special Holonomy,"
Oxford University Press, Oxford, 2000.
%
\bibitem{karigiannis}
S.~Karigiannis,
"Deformations of $G_2$ and $Spin(7)$ Structures on Manifolds,"
arXiv:math.DG/0301218.
%
\bibitem{complex}
J.~-H.~Lee and N.~C.~Leung,
``Instantons and branes in manifolds with vector cross product,''
arXiv:math.DG/0402044.
%
\bibitem{McLean}
R.~C.~McLean,
"Defomations of calibrated submanifolds,"
Commun.\ Anal.\ Geom.\ {\bf 6} (1998) 705.
%
\bibitem{Nekrasov:2004vv}
  N.~Nekrasov,
  ``A la recherche de la m-theorie perdue. Z theory: Casing m/f theory,''
  arXiv:hep-th/0412021.
%
\bibitem{Witten:1988xj}
  E.~Witten,
  ``Topological Sigma Models,''
  Commun.\ Math.\ Phys.\  {\bf 118} (1988) 411.
%
\bibitem{Witten:1991zz}
  E.~Witten,
  ``Mirror manifolds and topological field theory,''
  arXiv:hep-th/9112056.
%

  
\end{thebibliography}
\end{document}